\documentclass{appolb}
 \usepackage[dvips]{graphics}
  \usepackage{epsfig,epsf,amsmath,amssymb,bm}
   \graphicspath{{./figs-piff/}}

\begin{document}

\title{\hfill \hbox{RUB-TPII-07/09}\vspace*{5mm}\\
        Pion form factor in the QCD sum-rule approach\\
        with nonlocal condensates
       \thanks{Presented by the first author at the International Meeting
       ``Excited QCD'', February 8--14, 2009,
               Zakopane (Poland)}}

\author{Alexander~P.~Bakulev and A.~V.~Pimikov
 \address{Bogoliubov Laboratory of Theoretical Physics, JINR \\
          Dubna 141980, Russia\\
          E-mail: bakulev@theor.jinr.ru}
\and N.~G.~Stefanis
 \address{Institut f\"{u}r Theoretische Physik II,
          Ruhr-Universit\"{a}t Bochum\\
          D-44780 Bochum, Germany\\
          E-mail: stefanis@tp2.ruhr-uni-bochum.de\\
          and\\
          Bogoliubov Laboratory of Theoretical Physics, JINR \\
          Dubna 141980, Russia\\
          E-mail: stefanis@theor.jinr.ru
          }
}
\maketitle
\begin{abstract}
We present results of a calculation of the electromagnetic pion
form factor within the framework of QCD Sum Rules with
nonlocal condensates, using a perturbative spectral density
which includes $O(\alpha_s)$ contributions.
\end{abstract}

\markboth{\large \sl \underline{A.~P.~Bakulev et al.}
          \hspace*{2cm} Excited QCD, 2009}
         {\large \sl \hspace*{1cm} Pion form factor in the QCD SRs with NLCs}

\section{Introduction}
\label{sec:intro}
The pion---its distribution amplitude and form factor---stands tall
as a role model for the modern description of hadrons in terms of
quarks and gluons within QCD.
At high momenta $Q^2$, the pion form factor can be written as a
convolution
$F_\pi (Q^2)=\varphi_\pi^\text{out} \otimes T(Q^2) \otimes \varphi_\pi^\text{in}$
on account of the factorization theorem,
where the symbol $\otimes$ means
an integration over the longitudinal momenta of the quark and
antiquark in the pion state factorized at some scale $\mu^2$.
All binding effects due to the nonperturbative color dynamics are
absorbed into the pion distribution amplitudes $\varphi_\pi^\text{in}$
for the incoming and $\varphi_\pi^\text{out}$ for the outgoing pion.
However, at low and intermediate momenta $Q^2$
this factorization procedure
becomes inapplicable because the long-range interactions cannot be
separated into factorizing pion distribution amplitudes $\varphi_\pi$,
as above, so that the convolution approach is not very reliable.

In a recent paper \cite{BPS09}, we have proposed a different
theoretical framework for the calculation of the pion form factor,
which is based on QCD sum rules with nonlocal condensates (NLC)
\cite{MR86,MR92,BR91}, employing an Axial-Axial-Vector (AAV)
correlator.
Suffice it to say here that our scheme differs from previous ones
in several respects that will be considered in the next section.
A prolegomenon of the results we derived (see Sec.\
\ref{sec:predict}): the proposed method yields predictions for the
spacelike pion form factor that compare well with the data in that
momentum region which is currently accessible to measurements,
covering also the range of momenta to be probed by the 12 GeV upgraded
CEBAF accelerator at JLab.
Our conclusions are summarized in Sec.\ \ref{sec:concl}.

\section{Three-point QCD sum rule for the pion form factor}
\label{sec:theory}
Let us make here a second, more detailed pass, on some of the topics
briefly mentioned in the Introduction, recalling the AAV correlator
\begin{equation}
  \int\!\!\!\!\int\!\!d^4x\,d^4y\,e^{i(qx-p_2y)}
  \langle 0|T\!\left[J^{+}_{5\beta}(y) J^{\mu}(x) J_{5\alpha}(0)
               \right]\!
          |0
  \rangle\,,
\label{eq:Corr.JJJ}
\end{equation}
where $q$ denotes the photon momentum ($q^2=-Q^2$) and
$p_2$ is the outgoing pion momentum.
The quantities
$
 J^{\mu}(x)
=
 e_u\,\overline{u}(x)\gamma^\mu u(x)+e_d\overline{d}(x)\gamma^\mu d(x)
$
and
$J_{5\alpha}(x)= \overline{d}(x)\gamma_5\gamma_\alpha u(x)$,
$J^{+}_{5\beta}(x)= \overline{u}(x)\gamma_5\gamma_\beta d(x)$
are the electromagnetic current and the axial-vector currents,
respectively, where $e_u=2/3$ and $e_d=-1/3$ stand for the
electric charges of the $u$ and the $d$ quarks.
Referring for further details to \cite{NR82,IS82} for the case of
local condensates and to \cite{BR91} for nonlocal ones, we proceed
by writing down the sum rule we will employ:
\begin{eqnarray}
\label{eq:ffQCDSR}
  f_{\pi}^2\,F_{\pi}(Q^2)
& = &
  \int\limits_{0}^{s_0}\!\!\int\limits_{0}^{s_0}\!ds_1\,ds_2\
           \rho_3(s_1, s_2, Q^2)\,
           e^{-(s_1+s_2)/M^2}
\nonumber \\
&&     + \Phi_\text{G}(Q^2,M^2)
       + \Phi_{\langle\bar{q}q\rangle}(Q^2,M^2)\,.
\end{eqnarray}
Note that the quark-condensate contribution
\begin{eqnarray}
\label{eqquark-conden}
\Phi_{\langle\bar{q}q\rangle}(Q^2,M^2)
  =
     \Phi_\text{4Q}(Q^2,M^2)
    +\Phi_\text{2V}(Q^2,M^2)
    +\Phi_{\bar qAq}(Q^2,M^2)
\end{eqnarray}
contains the four-quark condensate (4Q), the bilocal vector-quark
condensate (2V), and the antiquark-gluon-quark condensate
($\bar q Aq$), while the term $\Phi_\text{G}(Q^2,M^2)$ represents
the gluon-condensate contribution to the sum rule.
The crucial quantity in the sum rule is the three-point spectral
density
\begin{eqnarray}
 \label{eq:SpDen.pert}
  \rho^{(1)}_3(s_1, s_2, Q^2)
  &=& \left[\rho_3^{(0)}(s_1, s_2, Q^2)
        + \frac{\alpha_s(Q^2)}{4\pi}\,
           \Delta\rho_3^{(1)}(s_1, s_2, Q^2)
    \right]\, .
\end{eqnarray}

The leading-order spectral density has been calculated long ago
\cite{NR82,IS82}, whereas the analogous next-to-leading order (NLO)
version
$\Delta\rho_3^{(1)} (s_1, s_2, Q^2)$
has been derived recently in \cite{BO04}.
The contribution from higher resonances is usually taken into account
in the form
\begin{equation}
 \label{eq:HR}
  \rho_\text{HR}(s_1, s_2)
  =  \left[1-\theta(s_1<s_0)\theta(s_2<s_0)\right]\,
      \rho_3(s_1, s_2, Q^2)
\end{equation}
and contains the continuum threshold parameter $s_0$.
In the investigation \cite{BPS09}, reported upon here, we use in
the perturbative spectral density a version of the running coupling
that avoids Landau singularities by construction (see for reviews in
\cite{SS97-06,AB08,Ste09}).
At the one-loop level, one has \cite{SS97-06}
\begin{eqnarray}
 \label{eq:alphaS}
  \alpha_s(Q^2)
   &=&\frac{4\pi}{b_0}
       \left(\frac{1}{\ln(Q^2/\Lambda_{\text{QCD}}^2)}
           - \frac{\Lambda_{\text{QCD}}^2}{Q^2-\Lambda_{\text{QCD}}^2}
                \right)
\end{eqnarray}
with $b_0=9$ and $\Lambda_\text{QCD}=300$~MeV.

The key elements of our analysis are these:
(i) All interquark distances in the quark-gluon-antiquark
condensate are nonlocal and the nonlocality is parameterized via
the quark-virtuality parameter $\lambda_q^2$ \cite{MR86} with the
value 0.4~GeV$^2$.
(ii) A modified Gaussian model for the nonlocal condensate is used,
recently proposed in \cite{BP06}, and the prediction for $F_\pi$ is
compared with the results derived in \cite{BPSS04}, obtained by
using the minimal Gaussian model \cite{MR92,BM98,BMS01},
and from other theoretical models~\cite{BT07,GR08} as well.
The virtue of the modified NLC model is that it helps minimizing the
transversality violation of the two-point correlator of vector
currents.
(iii) A spectral density is used that includes terms of $O(\alpha_s)$,
i.e., NLO perturbative contributions.
Moreover, the coupling entering the spectral density is analytic,
as stated above, so that the calculation of the pion form factor is
not influenced by the Landau pole.
It was shown in \cite{SSK99,BPSS04} that the Landau pole can obscure
the predictions for the pion form factor even at momentum values much
larger than $\Lambda_\text{QCD}$.

\section{Predictions}
\label{sec:predict}
\begin{figure}[t]
 \centerline{\includegraphics[width=0.60\textwidth]{
   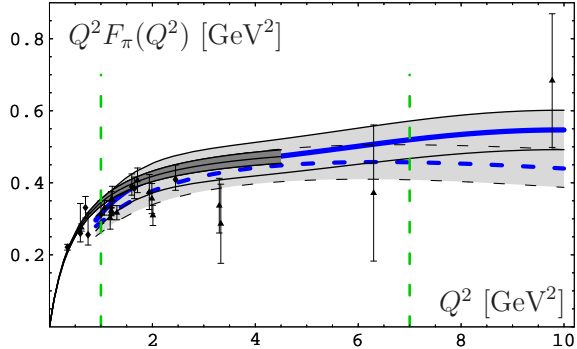}}
   \caption{\label{fig:main-result}\footnotesize
   Scaled pion form factor $Q^2 F_{\pi}(Q^2)$ for the minimal (dashed
   blue lines) and the improved (solid blue lines) NLC model using
   $\lambda_q^2=0.4$~GeV$^2$ in comparison with experimental data of
   the Cornell \protect\cite{FFPI74-78} (triangles)
   and the JLab Collaboration \cite{JLab08II} (diamonds).
   The shaded band within dashed lines shows the minimal model,
   whereas the variance of the improved model is indicated by solid
   lines.
   The lattice result of \cite{Brommel06} is also shown as a dark-grey
   strip.
   The two broken vertical lines mark the region, where the influence
   of the particular Gaussian NLC model used is not severe.
   }
\end{figure}

Using the presented scheme, we obtain predictions for the pion form
factor \cite{BPS09}, which are shown in Fig.\ \ref{fig:main-result}.
Our results are presented in the form of shaded bands, the aim being
to include the inherent theoretical uncertainties of the method.
The band contained within the solid lines gives the predictions we
obtained with the improved Gaussian NLC model, whereas the
corresponding findings from the minimal NLC model are shown in the
band with the dashed boundaries.
The central curves in each band
(illustrated by a thick solid and a thick dashed line)
are interpolation formulas,
whose explicit form can be found in \cite{BPS09}.
For the understanding of these predictions it is instructive to
remark that our method provides results that remain valid---though
with reduced accuracy---even at higher values of the momentum up to
$Q^2 \approx 10$~GeV$^2$.

Let us now discuss some technical details.
The value of $s_0(Q^2)$ at a given value of $Q^2$ is determined by
demanding minimal sensitivity of $F_{\pi}(M^2,s_0)$ on the Borel
parameter $M^2$ in the fiducial interval of the SR.
These intervals
$M^2\in[M^2_{-}/2, M^2_{+}/2]$
are determined from the corresponding two-point NLC QCD SR and
turn out to be $M^2_{-}=1$~GeV$^2$,
$M^2_{+}=1.7$~GeV$^2$ for the minimal NLC model and
$M^2_{-}=1$~GeV$^2$,
$M^2_{+}=1.9$~GeV$^2$ for the improved NLC model, while the
associated values of the pion decay constant read
$f_\pi=0.137$~GeV$^2$ and $f_\pi=0.142$~GeV$^2$,
respectively.
Notice that the value of the Borel parameter $M^2$ in the
three-point SR roughly corresponds to the Borel parameter in the
two-point SR, having, however, twice its magnitude.
A stable window for the Borel parameter is obtained for thresholds
in the range between 0.65 and 0.85~GeV$^2$ \cite{BPS09}.
As a rule, the higher the value of $s_0$, the larger the
form factor because the perturbative input increases.
The sensitivity of the obtained form-factor predictions on the
particular NLC model employed is rather weak in the region of
momenta delimited in Fig.\ \ref{fig:main-result} by two vertical
broken lines.
We close this discussion by remarking that the overall agreement
between the obtained predictions and the available experimental data
from~\cite{FFPI74-78,JLab08II} is rather good.
Moreover, they comply with the recent lattice calculation of
\cite{Brommel06}, which is shown in the same figure in terms of
a dark-grey strip,
bounded from high-$Q^2$ values at $Q^2\simeq4$~GeV$^2$.

In Fig.\ 2, we compare our predictions with other theoretical results,
derived from the AdS/QCD correspondence (so-called holographic QCD),
and some other models (including also the experimental data).
As in the previous figure, the shaded bands show our predictions for
the minimal (dashed lines) and the improved NLC model (solid lines).
The dashed (red) line at low-$Q^2$ (terminating at about 4~GeV$^2$)
represents the prediction derived in \cite{NR82} from QCD SR,
while the thicker broken line below all other curves is the result
of a calculation \cite{BLM07} based on Local Duality QCD SR.
The predictions from AdS/QCD are denoted by the upper short-dashed
(green) line---soft-wall model \cite{BT07}---and the penultimate
dash-dot-dotted (green) line, obtained with a Hirn--Sanz-type holographic
model \cite{GR08}.
The LD result of~\cite{BLM07} is shown as a dash-dotted (red) line.

\begin{figure}[h]
 \centerline{\includegraphics[width=0.60\textwidth]{
   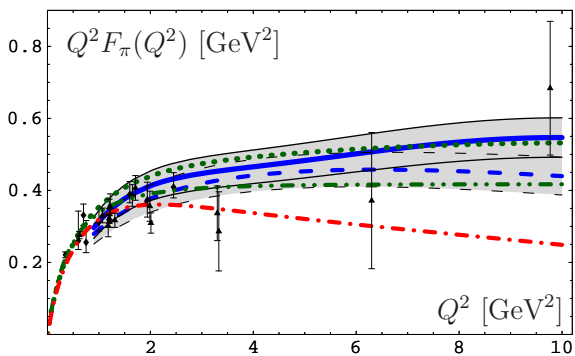}}
   \caption{\label{fig:main-result.2}\footnotesize
   Scaled pion form factor $Q^2 F_{\pi}(Q^2)$ within a band including
   uncertainties for the minimal (dashed lines) and the improved (solid
   lines) NLC model, using in both cases $\lambda_q^2=0.4$~GeV$^2$.
   The experimental data are as in Fig.\ \ref{fig:main-result}.
   The predictions from AdS/QCD are: upper short-dashed (green)
   line---soft-wall model \cite{BT07}; dash-dot-dotted (green)
   line---Hirn--Sanz-type holographic model \cite{GR08}.
   The dash-dotted (red) line corresponds to the LD result of~\cite{BLM07}.}
\end{figure}

\section{Conclusions}
\label{sec:concl}
Here we have studied a three-point AAV correlator within the QCD
sum-rule approach with nonlocal condensates in order to obtain
predictions for the spacelike pion form factor pertaining to that
momentum region accessible to experiment at present and in the
near future.
The full-fledged analysis can be found in \cite{BPS09},
where we also included in our discussion
the so-called Local-Duality approach~\cite{NR82}.

The principal ingredients of our approach are a spectral density that
includes $O(\alpha_s)$ corrections, whereas the coupling
used has an analytic structure without Landau singularities, and an
improved Gaussian ansatz for the nonlocal condensate.
Our findings are supported by both the existing experimental data
and also a recent lattice calculation in the momentum range up to
approximately 10~GeV$^2$.

\section*{Acknowledgements}
This work was supported in part by
the Russian Foundation for Fundamental Research,
grants No.\ ü~07-02-91557, 08-01-00686, and 09-02-01149,
the BRFBR--JINR Cooperation Programme,
contract No.\ F08D-001,
the Deutsche Forschungsgemeinschaft
(Project DFG 436 RUS 113/881/0-1),
and
the Heisenberg--Landau Programme under grant 2009.
A.~V.~P. acknowledges support from the Programme
``Development of Scientific Potential in Higher Schools''
(projects 2.2.1.1/1483, 2.1.1/1539, and 2.2.2.3/8111).

\end{document}